\documentclass[
    11pt,
    aip,
    rsi,
    reprint,
    notitlepage,
    superscriptaddress]{revtex4-2}

\usepackage{amsmath,amssymb,amsfonts} % standard AMS packages
\usepackage{empheq}
\usepackage{physics}

\usepackage[normalem]{ulem}

\usepackage{microtype}
\usepackage[svgnames,dvipsnames]{xcolor}
\usepackage{graphics}
\definecolor{NewBlue}{rgb}{0.1, 0.1, 0.7}
\definecolor{NewRed}{rgb}{0.7, 0.1, 0.1}
\usepackage[colorlinks,
    linkcolor=Maroon,
    citecolor=NewBlue,
    urlcolor=ForestGreen]{hyperref}

\usepackage[capitalise]{cleveref}

\usepackage[
    exponent-product=\cdot,
    range-phrase=\text{--},
    range-units=single,
    separate-uncertainty,
]{siunitx}[=v2]

\newcommand{\LigoMIT}{LIGO Laboratory, Massachusetts Institute of Technology, Cambridge, MA 02139}
\newcommand{\MechMIT}{Department of Mechanical Engineering, Massachusetts Institute of Technology, Cambridge, MA 02139}

\renewcommand{\t}[1]{\mathrm{#1}}

\begin{document}

\title{Temperature stabilization of a lab space at $\SI{10}{mK}$-level over a day}

\author{Dylan Fife}
\email{dsfife@mit.edu}
\affiliation{\MechMIT}

\author{Dong-Chel Shin}
\affiliation{\MechMIT}

\author{Vivishek Sudhir}
\affiliation{\MechMIT}
\affiliation{\LigoMIT}

\date{\today}

\begin{abstract} 
Temperature fluctuations over long time scales ($\gtrsim 1\,\t{h}$) are an insidious problem for
precision measurements. In optical laboratories, the primary effect of temperature fluctuations is 
drifts in optical circuits over spatial scales of a few meters
and temporal scales extending beyond a few minutes. 
We present a lab-scale environment temperature control system approaching $10\, \t{mK}$-level 
temperature instability across a lab for integration times above an hour and extending to a few days. This is achieved by 
passive isolation of the laboratory space from the building walls using a circulating air gap and 
an active control system feeding back to heating coils at the outlet of the laboratory HVAC unit. The
latter achieves 20 dB suppression of temperature fluctuations across the lab, approaching the limit set by 
statistical coherence of the temperature field.

\end{abstract}

\maketitle

\section{Introduction}

Long-term temperature drifts have always been the bane of precision measurements, stretching back to
pendulum-based timekeeping \cite{Jack29,Math04}, to contemporary atomic 
clocks \cite{Middle12,BoolYe14,Enzer17,Burt21,AeppYe24}. 
In experiments of this kind, the temperature of the critical subsystem can be 
stabilized \cite{Micali10,Sugi11,Middle12,Vicar19}.
In contrast, in complex optical experiments spread over a laboratory space, 
drifts in temperature need to be compensated
throughout the laboratory volume. To get a sense of the thermal time scale involved, notice that a 
3/16 inch thick, 4 foot $\times$ 10 foot steel top of an optical table acts like a thermal low-pass filter with 
an approximate cutoff frequency of $10^{-3}\, \t{Hz}$ (obtained from a simple lumped element heat conduction
model \cite{forsberg_chapter_2021}). 
So, for experiments running for more than 1000 seconds, one would expect that drifts in temperature 
of the optical table can produce drifts in the alignment of optical circuits built on it. 
In the worst case, $N$ mirrors each of transverse extension $\ell$, spread over an area $L^2$, will cause a beam displacement
of order $\delta x \approx (N L/\ell) (\partial \ell/\partial T) \delta T$, due to temperature fluctuation $\delta T$ coupling to the mirror
mounts through their thermal expansion coefficient $\partial \ell/\partial T \approx 10^{-5}\, \t{K}^{-1}$. 
For a typically complex optical circuit of $100$ $1"$ mirrors over an area of $4\, \t{m}^2$, this is a worst case drift
of $\delta x \approx (100\, \t{cm}) (\delta T/1\, \t{K})$.
% Extrapolating from the laser noise collected in this work, require thermal fluctuations of approximately 0.1 mK in the worst case (to get noise below $1 \t{prad}/\sqrt{\t{Hz}}$). In a system that is not so sensitive to thermal fluctuations, this condition may be relaxed.

This work demonstrates temperature stabilization of a laboratory-scale space
at the $10\, \t{mK}$ level over a time scale of a few days. 
Significantly better temperature stability has been achieved in volumes more than a million times 
smaller \cite{Micali10,Middle12,BoolYe14} by using a sequence of temperature shields and 
local thermo-electric control \cite{kocjan_temperature_2018, TECbox_grad}; similar performance has also been reported 
in meter-scale fully enclosed boxes \cite{Egan11}. 
At the scale of a room, air-conditioning alone is insufficient as that usually targets human
comfort and does not usually achieve stability better than a few degrees \cite{ghahramani_energy_2016}. 
A solution that has been pursued for precision physics experiments requiring temperature stability over
long times is to situate the experiment in a room-scale volume with a large thermal mass --- 
either underground \cite{Roll64}, 
or inside a mountain \cite{brack}, which can achieve sub-mK stability \cite{krishnan_generating_2020}. 
These solutions are essentially a large thermal enclosure that is unfit to be used as a general lab space, and they do not generalize beyond specific geographic locations.
Indeed, published reports from metrology and standards laboratories around the world \cite{Braud92,nist05,Lass11}
show that design and active control can achieve temperature stability around $100\, \t{mK}$; we demonstrate more than 
an order of magnitude improvement.

\section{Setup and methods}

The laboratory room of interest is situated in the basement of the historic main block of the 
Massachusetts Institute of Technology (MIT). The building, constructed in the early 20th century, has
walls facing the outside environment that are several feet thick; however, thermal insulation
is compromised by several factors: the presence of a window in that wall, 
two other walls lacking in thickness, the final wall needing to have a door opening into a corridor,
and the fact that MIT's central air handling unit (AHU) varies significantly in air temperature over a day.

\begin{figure}
    \includegraphics[width=0.9\columnwidth]{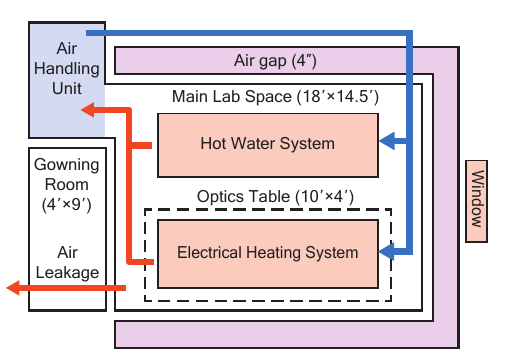}
    \caption{The lab space layout and the airflow within the space. Red arrows represent the flow of heated air through the lab space, whereas blue arrows represent the flow of cold air through the lab ductwork. The purple air gap is exposed to the temperature fluctuations from the external environment as well as the heated air in the lab and should be at some intermediate temperature.}
    \label{fig:airflow}
\end{figure}

Our lab space was designed and renovated for greater temperature stability while simultaneously 
meeting the competing requirements of reducing acoustic noise and maintaining a cleanroom environment 
(approximately Class 1000).
These requirements are met by a sequence of design choices. 
The laboratory space is constructed as a room-within-a-room: a few-inch air gap separates the walls of
the lab environment from the walls of the building on three sides (see \cref{fig:airflow}). 
Of the other surfaces, the one remaining wall opens into a gowning area that partly 
insulates the lab from the corridor; the ground is anchored to the thermal mass of the 
several-meter-thick foundation of the MIT main building; and the false ceiling sits below an air gap of
2 feet and contains air handling ducts.
The ducts serve air from a custom AHU, which contains a refrigeration system. 
The cooled air output from it is split in two (see \cref{fig:airflow}): 
one part passes through a heater driven by MIT's central hot water system and is then fed through HEPA
filters onto the part of the lab surrounding the central optical table; the other part passes through 
electrical coils behind separate HEPA filters over the optical table. 
(To reduce acoustic noise, careful attention was paid to the geometry of the ducts and air flux through them.)
Air from the lab space is recirculated 
back to the AHU via the air gaps on three sides of the lab. Effectively, this arrangement
produces a ``Russian doll'' thermal insulation of the optical table enclosure: 
first from the constant recirculation of air behind the lab walls, then from the air heated by the
central heaters, and finally, the air falling on the optical table can be heated by the user via 
an active control system. (The hot water system keeps the human-occupied area of the lab warm 
when active control of the temperature is off.)

Assuming that the temperature in the optical enclosure is mainly dictated by the temperature of the
air flowing into it, we actively stabilize the temperature of this air. A single thermistor (NTE 02-N102-1) 
in a four-wire configuration is polled at 5 Hz, and its signal is fed back to the electric heaters 
via a temperature controller (Stanford Research Systems PTC10). A second thermistor --- read out by the same
controller --- is used as an out-of-loop probe. 
The time series of temperature readings from both probes were recorded in various experiments, to be
detailed below, with no human occupation of the lab space. Power spectral densities (PSD) of the long-term
temperature time series were estimated using the Welch method \cite{Welch} using Hamming windows whose 
length spanned 4-14.5 hours ($\sim 2^{16}-2^{18}$ points) depending on the
length of the experimental run, an overlap of $50\%$, and linear 
detrending to avoid bias in the lowest frequency bins of the PSD estimates. In order to quantify the effect of long-term
drifts, we also investigate the modified Allan deviation of the 
temperature data \cite{riley_handbook_2008}.

\begin{figure}[t!]
    \includegraphics[width=0.95\columnwidth]{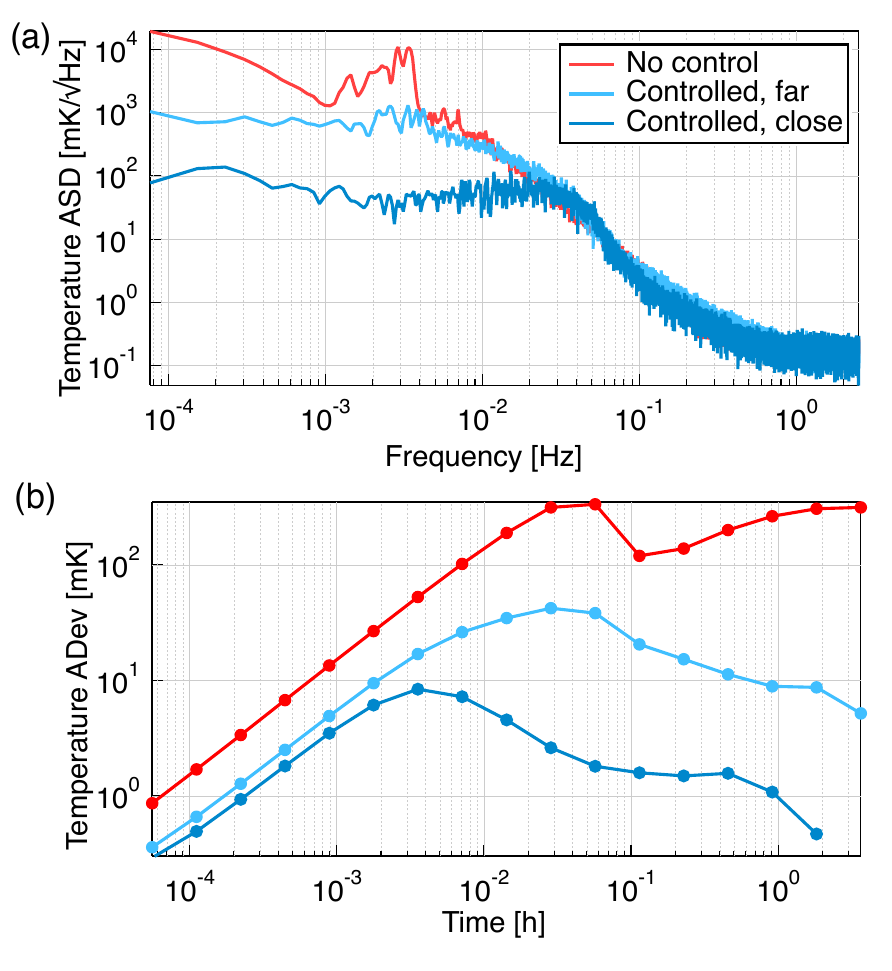}
    \caption{\label{fig:temperature}
    (a) shows the ASD of the free-running temperature (red), and when the temperature is controlled (blue);
    dark blue shows the temperature measured a few mm away from the sensor used for control, while
    light blue shows the temperature measured $2.35\, \t{m}$ away from the sensor.
    (b) shows the modified Allan deviation of the same time series underlying the data in panel (a); colors are
    identical to panel (a).}
\end{figure}

\section{Results}

To ascertain the performance and effect of the passive and active temperature stability of the 
laboratory space, several measurements were performed.
First, we measured the temperature fluctuations with and without the active control system, with the two sensors
placed next to each other. 
Second, the sensors were placed across the room from each other to quantify the effect of control over
large spatial scales.
Third, a laser beam was bounced across the length of the optical table to measure the effect of
temperature stability on the geometric stability of the beam path.
In the following, we detail the results of these investigations.

In the first experiment, the controller sent a 
constant reference voltage to the heater, but the feedback loop was not closed. In essence, we measure the
passive stability of temperature in the lab space without active feedback control.
Data is recorded continually over three days, and its amplitude spectral density (ASD) and modified Allan deviation 
are shown in red in \cref{fig:temperature}.
The peaks in the temperature spectrum around frequency $f\approx (5\, \t{min})^{-1}$ corresponds
to the typical time over which the AHU takes in air from the outside.
However, passive isolation already suffices to keep the temperature stable to within $\sim 200\, \t{mK}$ as apparent
from the Allan deviation.

Next, active feedback based on a single thermistor is turned on, and a second (out-of-loop) thermistor is
placed in its vicinity as a witness of temperature. 
Given the hours-long response time of the temperature in the room, we employ
the relay tuning feature of the controller (which sends a series of steps into the heater and observes
the response on the thermistor to tune a PID filter). 
The dark blue lines in \cref{fig:temperature} show the data from such an experiment.
Apparently, the feedback loop has a 
unity-gain frequency of $\sim 2\cdot 10^{-2}\, \t{Hz}$, which suffices to affect a 100-fold reduction
in the ASD. As evident from the Allan deviation, active control succeeds in attaining a stability less
than $1\, \t{mK}$ over three days \emph{in the vicinity of the sensing thermistor}.

Clearly, temperature fluctuations faster than $10^{-2}\, \t{Hz} \approx 100\, \t{s}$ are not suppressed
by active control. This is largely due to the delay in the feedback loop arising from 
the slow response of the heaters and the concomitant limitation in closed-loop stability that curtails
bandwidth. In any case, smaller parts of an experiment can be isolated from such high-frequency temperature
fluctuations by thermalizing them with a large thermal mass, which acts like a low-pass filter. 
The thermistor, by comparison, has a lower thermal mass than most optical components. 
As such, any optical component is unlikely to be susceptible to temperature fluctuating faster
than $100\, \t{s}$.

In a second experiment, to quantify the efficacy of the temperature control over the spatial scale of the lab, 
we move the out-of-loop sensor to a distance of 2.35 m from the in-loop sensor.
The light blue traces in \cref{fig:temperature} show the results. 
There are noticeable gains in temperature fluctuations around the peak, but the temperature across the lab is more noisy compared to the temperature in the vicinity of the in-loop sensor. Even so, within about an hour, 
the temperature across the lab is stable to within 10 mK, as shown by the Allan deviation.

\begin{figure}[t!]
    \includegraphics[width=0.9\columnwidth]{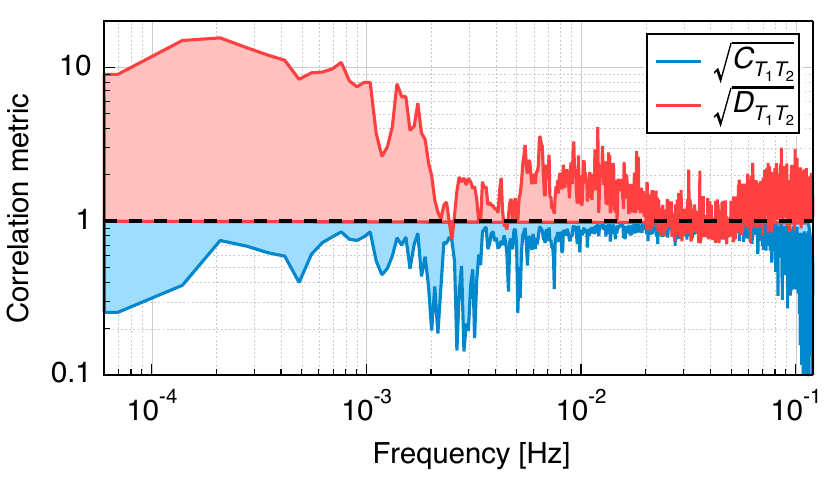}
    \caption{Blue shows the coherence $C_{T_1 T_2}$ between the temperatures measured by the 
    two sensors when they are separated by 2.35 m. The two temperatures lack coherence at the lowest frequencies. 
    Red shows the figure of merit $D_{T_1 T_2}$
    defined in \cref{eq:D}. The excess value of $D$ over unity is consistent with the degree of
    incoherence.
    \label{fig:coherence}}
\end{figure}

We expect that much of the residual noise across the lab is not coherent across the lab space, 
and so our control system, which only has one 
degree of freedom, cannot compensate for uneven temperature gradients. 
To ascertain the limitation posed by the incoherence of the temperature field across the lab, 
consider the temperature fluctuations $T_1$ and $T_2$ at any two locations. The power spectral density
of their difference $S_{T_1 - T_2}$ is zero if and only if the two random processes have identical
trajectories in any given sample. This quantity is, however lower-bounded by the 
(magnitude of the) coherence $C_{T_1 T_2} \equiv 
\abs{S_{T_1 T_2}}/\sqrt{S_{T_1} S_{T_2}}$, as follows:
\begin{equation}
\begin{split}
	S_{T_1 - T_2} &= S_{T_1} + S_{T_2} - 2 \Re S_{T_1 T_2} \\
		&\geq 2\sqrt{S_{T_1}S_{T_2}} - 2 \abs{S_{T_1 T_2}} \cos(\arg S_{T_1 T_2}) \\
		&\geq 2\sqrt{S_{T_1}S_{T_2}} (1 - C_{T_1 T_2}).
\end{split}
\end{equation}
Thus, $S_{T_1 - T_2} = 0$ if and only if the coherence attains its maximum value 
$C_{T_1 T_2} = 1$, i.e. iff. the temperature field is perfectly coherent
across the two locations. (The fact that $C_{T_1 T_2} \leq 1$ follows from the Cauchy-Schwarz inequality.)
Alternatively, if the temperature across the two locations is not perfectly coherent (i.e. $C_{T_1 T_2}\neq 1$), 
any attempt to control one based on a measurement of the other will not be perfect, 
and the lack of imperfection due to this effect can be quantified by the figure of merit, 
\begin{equation}\label{eq:D}
	D_{T_1 T_2} \equiv \frac{S_{T_1 - T_2}}{2\sqrt{S_{T_1}S_{T_2}} (1 - C_{T_1 T_2})} \geq 1.
\end{equation}
\Cref{fig:coherence} shows this figure of merit in red, and the coherence $C_{T_1 T_2}$ in blue. 
$D_{T_1 T_2}$ was in excess above its realistic lower bound of unity by a factor of 10, which is
commensurate with the drop in coherence (from its upper bound of unity) by roughly the same factor.
This shows that the current temperature control system approaches the limit set by the statistical coherence of the
temperature field, with only marginal improvements possible by tuning the controller.

\begin{figure}[t!]
    \includegraphics[width=0.9\columnwidth]{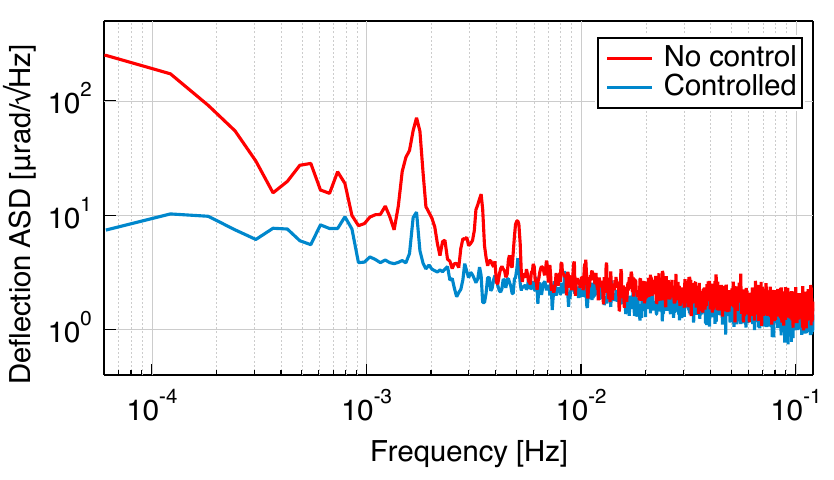}
    \caption{\label{fig:deflection}
    ASD of the angular deflection of a laser beam traversing a 6 m path along an optical table inside the
    lab space. Active control of the temperature leads to an order-of-magnitude reduction in drift down to
    $10\, \mu\t{rad/\sqrt{Hz}}$.}
\end{figure}

Our ultimate interest is in the effect of temperature drift as it pertains to drifts in optical
circuits in the lab. To characterize this effect, we measured the angular deflection of a laser beam
traversing the perimeter of our optical table. A beam was derived from a 1064 nm laser and
reflected off of 5 mirror mounts (Thorlabs KM100) around the table and measured using a quadrant photodetector (QPD). 
In order to cancel beam pointing noise intrinsic to the laser, a second beam was 
derived using a beam-splitter and measured on another QPD co-located with the first QPD after the beam traversed
the shortest possible path between the beam-splitter and the QPD. 
The subtracted photocurrent from the two QPDs (which largely eliminated the common pointing noise of the two beams) 
was used as a proxy of the angular deflection of the beam traversing the perimeter of the optical table.
\Cref{fig:deflection} depicts the resulting signal calibrated in angle units, with and without the temperature control.
It is apparent that stabilizing the temperature of the laboratory space noticeably decreases low-frequency drifts in
the laser beam's deflection.

\section{Conclusion}

We designed and commissioned a laboratory space for precision measurements that is temperature-stable at the 
$\sim 10\, \t{mK}$-level after an hour of averaging and over day-long time scales. 
This level of stability is about an order of magnitude improvement over laboratory scale temperature control 
demonstrated previously\cite{Braud92,nist05,Lass11}, and is achieved by a combination of passive isolation 
and active control of the entire lab space.
Active temperature control suppresses the passively isolated temperature by a 100-fold.
The achieved temperature stability manifests as an order of magnitude improvement in the angular drift of a 
laser beam traversing a $\sim 6\, \t{m}$ path around an optical table, at the level of $10\, \mu\t{rad/\sqrt{Hz}}$. 
Since the scheme presented here is independent of the specific use-case or geographical location of the lab,
it can be transplanted into the design of any precision measurement lab.

Importantly, the performance of our single degree of freedom controller approaches the limit set by
the statistical coherence of the temperature field being controlled. Thus, significant improvement 
can only be achieved by techniques such as sensor fusion of a network of temperature sensors and spatially
localized actuation.
\section*{Data Availability}
The data that support the findings of this study are available from the corresponding author upon reasonable request.

\section*{Acknowledgements}

The design and construction of the lab was made possible by the generous support of the MIT School of Engineering
and the Department of Mechanical Engineering. We also thank Pierre Lermusiaux, Jon Keller, Alireza Heidari, Doug Bell, 
Dorian Alba, William Mackie, Thomas Graffeo, and Alex Farrell.

\bibliography{thermal.bib}

\end{document}